\documentclass[preprint,fleqn]{aastex}
\usepackage{graphicx}
\newcommand{\EN}[1]{ \vec{e}_{#1} }
\newcommand{\ER}{\vec{e}_\varpi}
\newcommand{\EP}{\vec{e}_\phi}
\newcommand{\EZ}{\vec{e}_z}
\newcommand{\Div}{\vec{\nabla} \cdot }
\newcommand{\Grad}{\vec{\nabla} }
\newcommand{\EV}{\vec{E}}
\newcommand{\BV}{\vec{B}}
\newcommand{\VV}{\vec{v}}
\newcommand{\JV}{\vec{j}}
\newcommand{\FV}{\vec{F}}
\newcommand{\VXI}{\vec{\xi}}
\newcommand{\ALF}{Alfv$\acute{\rm{e}}$n }
\begin{document}
\title{ Propagation and Transmission of \ALF Waves
in Rotating Magnetars   }
%
%
\author{Yasufumi Kojima and Taishi Okita }
\affil{
Department of Physics,
Hiroshima University,
Higashi-Hiroshima 739-8526
Japan}
\email{kojima@theo.phys.sci.hiroshima-u.ac.jp}
\begin{abstract}
  We study the propagation and transmission of \ALF waves
in the context of cylindrical geometry.
This approximates the polar cap region of aligned pulsar
with strong magnetic fields. 
Non-propagating region appears in the presence of rotation.
The displacement current further prevents the low frequency 
modes from propagating near the stellar surface.
The transmission rates to the exterior through the surface
are calculated. The rates increase with the frequency and
the magnetic field strength. The transmission also depends
on the helicity states of the waves, but
the difference becomes small in the high frequency regime.
We also point out the possibility of the spin-up by 
outgoing wave emission in the low frequency regime,
if a certain condition holds.
\end{abstract}

\maketitle

\section{Introduction}
   Soft Gamma Repeaters (SGRs) and Anomalous X-ray Pulsars (AXPs) 
belong to a rare class among thousands of neutron stars.
Both objects are different from so far known neutron stars
in their radiation spectrum  and spin periods.
SGR-like outbursts were also discovered in AXPs \citep{gkw,ka},
and hence there are similarities between two peculiar 
neutron stars.  They are likely to be young and isolated, 
but have intense magnetic fields in the $ 10^{14}-10^{15}$G range.
See  \citet{ms1, kds} for their initial observations 
and also  \citet{w1} for recent reviews.
  Further evidence of the strong magnetic fields has 
recently come from spectral line feature\citep{isp}.
These objects are called as magnetars \citep{td1,td2,td3}.
The magnetar model has been able to explain
the peculiar observational properties. See  e.g.
\citet{t1, ltk} as recent reviews. 
The burst emission and non-thermal X-ray radiation
are supplied by the decay of strong magnetic fields.
Twisted magnetic fields relevant to  
the activities are transported from the core to the
surface by ambipolar diffusion in the magnetars model
\citep{tlk}.
%

 The \ALF waves are likely to be excited by sudden  disturbances
in various magnetized objects. Short duration bursts in SGRs may
be associated with starquakes  driven by magnetic stress,
since the similarity in statistical property
is reported between the SGRs and earthquakes
\citep{chen, go1, go2}.
   The \ALF waves are very interesting in their property 
and physical mechanism. The fluid displacement is 
perpendicular to both of the wave vector and the magnetic field,
and restoring force arises from the
magnetic tension of the field lines.
The plane \ALF waves of finite amplitude can propagate
at constant speed in a homogeneous incompressible medium
without any distortion of the waveform.
This  contrasts with sound waves, which may
steepen to form shocks because of the non-linearity.
As for the physical mechanism,
torsional \ALF waves can transport angular momentum, and
account for the spindown of rotating objects, such as
magnetized stars with convective envelopes,
interstellar clouds threaded by the Galactic magnetic fields
and so on. See e.g. \cite{shu} for details.

  We consider the  \ALF waves produced by shaking the magnetic
field lines in a rotating magnetar in this paper.
The origin of abrupt disturbances relevant to the bursts
is not addressed here, but the subsequent propagation and 
ejection to the exterior are examined.
The total energy in shorter duration bursts
is $\Delta E < 10^{41}$ ergs, and is a small fraction of
available magnetic energy $\sim 10^{-5} E_B$,
where the total magnetic energy $ E_B$ is estimated 
by the expected dipole field strength $B=10^{14} -10^{15}$G.
On the other hand, giant flares with $\Delta E \sim 10^{44}$ ergs
should involve more drastic change of magnetic field
configuration on a global scale \citep{io,w0}.
We here use the linearized perturbation equations,
which are applicable to less energetic events
$\Delta E /E_B  \propto  (\delta B/B )^2\ll  1  $, 
in the shorter duration bursts.
Our results of the wave propagation may not be applied to the 
giant flares, but there may be some similarities in nature
even for $ \delta B \sim  B$.
The ejection of \ALF waves to the exterior was already 
estimated so far, but our model is beyond a simple model.
We here explicitly calculate the transmission in a 
cylindrical geometry.
We also calculate the energy and angular momentum 
extracted by the waves through the stellar surface.
In section 2, we provide a  geometrically simplified model.
The propagation in the neutron-star
crust and transmission to the exterior can be examined
in a concrete way owing to the simplified model.
  The shear in the solid crust acts as
additional restoring force. Therefore, the shear-\ALF waves
may be more adequate for the name.
The transmission rate is numerically calculated for the waves
with higher frequency.  We discuss the implication of our 
results in section 3. 
%

\section{ \ALF Waves in Rotating Cylinder}
\subsection{Model}
%
In this paper, we study the propagation of the shear-\ALF waves 
through a neutron-star crust and the transmission to the exterior. 
The star is assumed to rotate around $z$-axis with 
a constant angular velocity $ \Omega $.
The magnetic field is uniform along the $z$-axis, i.e.
$ \BV = B_0 \EN{z}$, where $ B_0$ is a constant.
Our consideration is limited to a cylindrically symmetric 
slab region within the radius $\varpi _*$.
The magnetic fields are rather easily incorporated 
in the cylindrical model.
\cite{car86} studied the oscillation spectra of the
magnetized/unmagnetized cylindrical stars.
Their results show that the periods of the torsional and 
shear modes are in good agreement with those of spherically
symmetric unmagnetized star.
This suggests that the approximation is good for some modes.

The cylindrical approximation may be adequate for
the polar cap region and the interior for an aligned rotator. 
In this case, the radius $\varpi _*$ is 
given by the size of the polar cap,
$ \varpi _*  \le  R \sin \theta _p  
=  \sqrt{  R^3 \Omega /c } 
\approx 10^4 (T/1{\rm s})^{-1/2}  ~{\rm cm }$,
where $ R \approx 10^6  ~{\rm cm }$ is stellar radius
and $T $ is the rotational period.
The exterior magnetic field lines originated from this region
are extended to infinity. 
On the other hand, the field lines outside 
the polar cap become closed ones. The plasma along the closed 
lines corotates with the central star. In our cylindrical model,
we consider the shaking of open field lines only, which may 
lead to particle acceleration, magnetic  reconnection, wind,
and  eventually to the radiation.
Explicit treatment of the energy transfer to
the radiation is also beyond the scope of this paper.

The shear stress in the crust also acts as a restoring force,
and is coupled to the \ALF waves.
The solid crust ranges from $ z=- d $ to the surface $ z = 0 $.
The depth  $d$ of the outer crust is given by
$d  \approx 10^{5}~{\rm cm }$.
Our consideration is limited to thin outer crust only.
The slab geometry is valid there because  $d \ll R$.
Gravitational acceleration in this region is
almost constant and is approximated as $ -g \EN{z}$ ,
with $ g =10^{14} {\rm cm~s}^{-2} $.
The density distribution $\rho(z)$  is given by the
integration of hydrostatic equation
for neutron star crust ($|z| < d $), where 
degenerate electron pressure is dominated.
The analytic expression is available as
\citep{BBGP}
\begin{equation}
  \rho = 8.0 \times 10^1 \left[
(|z|/{\rm cm }) + 2.5 \times 10^{-4} (|z|/{\rm cm })^2
\right] ^{3/2} {\rm g~cm}^{-3}  .
\label{rhoz}
\end{equation}
This explicit form will be used for the numerical 
calculations in the subsequent sections.

\subsection{Wave equation}
We consider the propagation of shear-\ALF waves
based on the linearized perturbation theory.
In the deep interior, gravitational force is
so large that  the displacements to the 
vertical direction are not easily to be induced.
The  horizontal displacements of the disturbances  are likely
to be  dominated, and they are coupled with the \ALF waves.
  Some restrictions are imposed on the displacement vector
$\VXI$ of the materials in order to extract the waves 
coupled with shear and magnetic stress.  
The waves propagating along the $z$-axis are assumed to satisfy
$\xi_z =\Div \VXI =0$.
This means that the waves are transverse and are decoupled from
the compressional modes.
The Lagrangian perturbations of density and pressure are zero
because they are proportional to $ \Div \VXI $, i.e.
$   \Delta \rho /\rho   = - \Div \VXI  =0 ,$ 
$   \Delta p /p  = - (\partial \log p/ \partial \log \rho )_{ad}
 \Div \VXI =0 .$ 
  The general forms of the displacements 
can be expressed by the Fourier and  Bessel functions
with respect to time and cylindrical radius, respectively.
We further simplify the displacements by assuming
nodeless functions in cylindrically radial direction,
i.e. neglecting the radial structure. The regular form near
the  $z$-axis is simply given in the cylindrical coordinate
$(\varpi, \phi,z)$ as
\begin{equation}
\label{disp}
  \VXI \propto  ( \ER \pm i \EP ) \varpi ^{m-1}
  \xi _{\pm~ m~ \omega } (z) e^{-i (\omega t \mp m \phi)  },
\end{equation}
where $m$ is a positive integer, $m \ge 1$.
The mode for the upper sign 
in  eq.(\ref{disp}) is said to have positive helicity, 
whereas that for lower sign has negative helicity.
We may limit the Fourier mode to a certain frequency region,
using a general relation 
$ \xi _{\pm~m~-\omega }=\xi ^*  _{\mp~ m~ \omega }$,
where $^* $ means complex conjugate.
We define the frequency in corotating frame 
$ \sigma_{\pm} \equiv \omega \mp m \Omega $
for each helicity state $\xi _{\pm} $
and limit the range to $ \sigma_{\pm} \ge 0$.
From now on, we  will omit the suffixes ${m~ \omega}$
in order to avoid complicated notations.
The displacement  (\ref{disp}) satisfies
 $({\vec \nabla} \times \VXI )_z = 0 $
and is therefore  decoupled from the vorticity as it is desired.
The displacement corresponding to $m=1$ has clear meaning,
and can be written in the Cartesian coordinates as
$  \VXI =  ( \EN{x}  \pm i \EN{y} )
  \xi _{\pm}  e^{-i \omega t  }.$
This mode represents uniform motion in horizontal direction.
A slightly different treatment is necessary for
the axially symmetric perturbation $ m=0$.
The regular displacement satisfying above conditions
is  given as
$  \VXI =  \xi _\phi ( t, \varpi) \varpi\EP $,
which corresponds to the velocity perturbation
$  \delta \VV  =  (\partial \xi _\phi /\partial t ) \varpi\EP $
$ \equiv  \delta \Omega \varpi \EP $.
This kind of perturbation merely represents
impulsive jump of angular velocity as
$ \Omega \to \Omega + \delta \Omega$,
and is no longer  considered in  this paper.

The restoring forces for the modes are elastic shear stress 
$\delta S_i$ and electromagnetic force $\delta F_i$.
The linearized equation of motion can be written as
\begin{equation}
\rho \left(
  \frac{ \partial }{\partial t} \delta  v_i 
+ ( \VV \cdot \Grad ) \delta  v _i 
+  ( \delta \VV \cdot \Grad ) v _i \right)
= 
\delta S_i +  \delta  F_i ,
\label{eqom.eqn}
\end{equation}
where $\VV=\varpi \Omega \EP$,  and the relation between 
the displacement and velocity perturbation is
\begin{equation}
  \delta v _i =  \frac{ \partial }{\partial t} \xi _i
+  ( \VV \cdot \Grad )   \xi _i 
-  ( \VXI \cdot \Grad ) v _i . 
\end{equation}
The shear stress $ \delta S_i $  associated with deformation is
\begin{equation}
 \delta S _i = \nabla _j   
\left[  \mu \left(
\frac{\partial \xi^i}{\partial x^j} 
+
\frac{\partial \xi^j}{\partial x^i} 
\right) \right] ,
\end{equation}
with shear modulus 
$ \mu
  = 4.8 \times 10^{27} 
\left( \rho / 10^{11} {\rm g ~ cm}^{-3} \right)^{4/3}
~{\rm erg~ cm}^{-3}
$
\citep{BP}.
The electromagnetic force is given by 
\begin{equation}
 \delta \FV = 
\delta \rho_e  \EV + \rho_e \delta \EV
+  \frac{1}{c} ( \delta  \JV \times \BV  
+  \JV \times \delta \BV )  .
\label{emforce}
\end{equation}
The crust is a perfect conductor, so that
the electric fields for both unperturbed and perturbed states 
are induced by the material motion as
\begin{equation}
  \EV = - \frac{ 1 }{c} 
\left(  \VV \times \BV
\right) = - \frac{\varpi \Omega B_0 }{c} \ER,
\label{ee.bk}
\end{equation}
\begin{equation}
  \delta \EV = - \frac{ 1 }{c} 
\left( 
     \delta \VV \times \BV
 +   \VV \times \delta \BV
\right).
\label{froz.eqn}
\end{equation}
The electric charge and current for the unperturbed state are
\begin{equation}
\rho_e = -\frac{1}{2 \pi c} \Omega B_0,
~~~~~
\JV = \rho_e \VV .
\label{jc.bk}
\end{equation}
These quantities (\ref{ee.bk}) and (\ref{jc.bk})
satisfy the force balance
$ \rho_e \EV  + \JV \times \BV / c =0$.
These expressions are valid in the region smaller than 
light cylinder $ c/\Omega$  or actual stellar radius $ R$.
Otherwise, we would need better approximation
beyond the uniformly rotating cylindrical model.
Therefore, our present model may be adequate to
the behavior near the $z$-axis.
As will be confirmed by the explicit forms 
of $ \EV $ and $ \delta \EV$, 
the first term 
$ \delta \rho_e \EV ( = (\Div \delta \EV /4 \pi  )\EV  )$
 in eq.(\ref{emforce}) is proportional to
$ (\Omega \varpi / c )^2$ and  is smaller than other terms
near the $z$-axis, i.e. 
for the region $ \varpi ^2 < \xi/ \xi'' $.
We therefore neglect the first term $ \delta \rho_e \EV$
in the propagation of \ALF waves.
Eliminating $\delta \EV$ by eq.(\ref{froz.eqn}), 
the Lorentz force (\ref{emforce}) is reduced to
\begin{equation}
 \delta \FV =
  \frac{1}{c} ( \delta  \JV  - \rho_e \delta \VV ) \times \BV .
\end{equation}
The perturbation of electric current is determined by the 
Maxwell's equations:
\begin{equation}
\nabla  \times \delta \EV  
= -\frac{1}{c} \frac{ \partial }{ \partial t } \delta \BV ,
\label{ind.eqn}
\end{equation}
\begin{equation}
\nabla  \times \delta \BV  
= \frac{ 4 \pi }{ c  } \delta \JV 
+\frac{1}{c} \frac{ \partial }{ \partial t } \delta \EV .
\label{amp.eqn}
\end{equation}
From eqs.(\ref{froz.eqn}) and (\ref{ind.eqn}),
the perturbations of electro-magnetic fields are
expressed in  terms of the displacement vector as
\begin{equation}
\label{delB}
\delta \BV  
= B_0 \nabla_z \VXI =
 B_0 \varpi ^{m-1}  
\frac{d   \xi _{\pm} }{dz} ( \ER \pm i \EP ) 
e^{-i (\omega t \mp m \phi)  } ,
\end{equation}
\begin{equation}
\delta \EV  
=  \frac{ B_0 \varpi ^{m-1} }{c} 
\Big[
\mp (\omega  \mp m \Omega)  \xi _{\pm}
( \ER \pm i \EP )
+  \Omega \varpi 
\frac{d  \xi _{\pm} }{dz} 
\EZ
\Big] e^{-i (\omega t \mp m \phi)} .
\end{equation}
The stellar rotation induces the $z$-component of the 
perturbed electric field, which is important for
angular momentum transfer as discussed in section 2.6.
Using these expressions and
eliminating $\delta \JV$ by eq.(\ref{amp.eqn}), 
eq.(\ref{eqom.eqn}) is eventually reduced to
\begin{equation}
 \frac{ d }{ dz } 
\left( \mu + \frac{B_0 ^2}{4\pi }  \right)
\frac{ d }{ dz } \xi_{\pm }
+ V_\pm \xi_{\pm }  =0,
\label{base.eqn}
\end{equation}
where
\begin{equation}
V_{\pm}    = 
\left( \rho +\frac{ B _0 ^2  }{4\pi c^2} \right)
 (\omega \mp m  \Omega ) \left[ \omega \mp (m h -2) \Omega \right],
\end{equation}
\begin{equation}
 h = \frac{4\pi c^2 \rho }{ 4\pi c^2 \rho  + B _0 ^2 } .
\end{equation}
%

\subsection { Propagation }

  The propagation of \ALF waves is studied in different contents.
In a certain limit of parameters, our basic equation (\ref{base.eqn})
should be reduced to one previously studied. 
In the limit of non-rotating case, eq.(\ref{base.eqn})
is reduced to one found by \cite{BBGP}.
The term $V_{\pm} $ is independent of the helicity state
and is positive definite in this case.
Another interesting limit of eq.(\ref{base.eqn}) is obtained 
when both shear and relativistic effects are neglected.
This limit corresponds to $\mu \to 0$ and  
$c^2 \to \infty $. 
In this case, the term $V_{\pm}$ becomes negative
for a certain frequency region.
Hence, the modes are non-propagating (evanescent).
See e.g. \cite{chand}.
We will discuss the evanescent property below.

We consider the short-wavelength limit in order 
to see whether or not the waves propagate.
The local dispersion relation is obtained by
setting $ \xi _{\pm} \propto e^{ ik z }$ 
in eq.(\ref{base.eqn})
\begin{equation}
  -v^2 k^2 +
(\omega \mp m  \Omega )(\omega \mp (m h -2) \Omega )
=0,
\label{disp.eqn}
\end{equation}
where $v$ is the velocity defined as
\begin{equation}
  v =c 
\left(
\frac{ 4\pi \mu+ B _0 ^2}{
4\pi \rho c^2+ B _0 ^2}  \right)^{1/2}.
\end{equation}
In the limit of $\mu \to 0$ and $c^2 \to \infty$, 
the classical \ALF wave velocity is recovered, i.e.
$ v= B_0 /\sqrt{4\pi \rho }.$
The shear acts as the restoring force deep in
the crust unless $  B_0  \gg 2 \times 10^{14}$ G.
The displacement current contributes to additional inertia 
and is always important near the surface, where the 
propagation velocity becomes $c$.
Solving eq.(\ref{disp.eqn}) for the frequencies 
$ \sigma _{\pm} \equiv  \omega \mp m \Omega  \ge 0$, 
we have 
\begin{equation}
 \sigma _{\pm}  
 = \mp \left[ 1 + \frac{m}{2}(1-h) \right] \Omega 
 + \left\{ \left[ 1 + \frac{m}{2}(1-h) \right] ^2 \Omega^2 
 + v^2 k ^2 \right\}^{1/2} .
\end{equation}
We do not have to consider the modes with negative 
frequency $ \sigma _{\pm} < 0$ because of
the correspondence $ (\xi_{\pm } , \sigma  )$
$ \leftrightarrow (\xi_{\mp }^* , -\sigma  )$
as mentioned in section 2.2.
The difference between two helicity modes
is clear in weak magnetic field limit, i.e.
$ v k / \Omega  \to 0 $ and $ h   \to 1 $.
In this limit,
the frequencies in the corotating frame 
are given by
$ \sigma _{-}  \approx  2 \Omega $
and
$ \sigma _{+} \approx  v^2 k ^2/( 2 \Omega ) $.
The mode $ \xi_{-}$ with frequency $ \sigma _{-} $
represents an inertial wave due to Coriolis force.
On the other hand, the mode $ \xi_{+}$ with
smaller frequency $ \sigma _{+} ~(\ll \sigma_{-}) $
represents a drift wave.
Our problem is limited to cylindrical geometry,
but the similar oscillations are also possible in 
spherical geometry. See e.g. \cite{Levin}
for comparison in the non-relativistic limit, 
although careful treatment is necessary to convert 
results between different geometry.
  Like the \ALF waves in classical treatment,
the waves with low frequencies 
become non-propagating due to the Coriolis force.
The mode $\xi _{+} $ is always propagating as far as
$ \sigma_{+} \ge 0 $,
whereas the mode $\xi _{-} $ is non-propagating
for the frequency region
$ 0 \le \sigma_{-} /\Omega  \le  2 + m (1-h) $. 
The condition becomes
$ 0 \le \sigma_{-} /\Omega  \le  2  $,
when the relativistic effects are neglected.
The evanescent property in our problem
depends on the spatial position through $h(z)$,
which is calculated for the density distribution 
(\ref{rhoz}) and constant magnetic field strength $ B_0$. 
It is general that the function $h $ 
decreases from $h \sim  1 $ in deep interior 
to $ h  \sim 0$ near the surface with low density.
The evanescent region therefore prevails near the surface.
In Fig.1, we demonstrate the location of evanescent region 
using the local dispersion relation.
The critical frequency 
$ \sigma_{-} /\Omega = 2 + m (1-h) $  is shown.
The depth of the evanescent region  increases 
with azimuthal wave number $m$ as well as
the magnetic field strength $B_0$.
%

\begin{figure}[ht]
\centering
\includegraphics[scale=1.0]{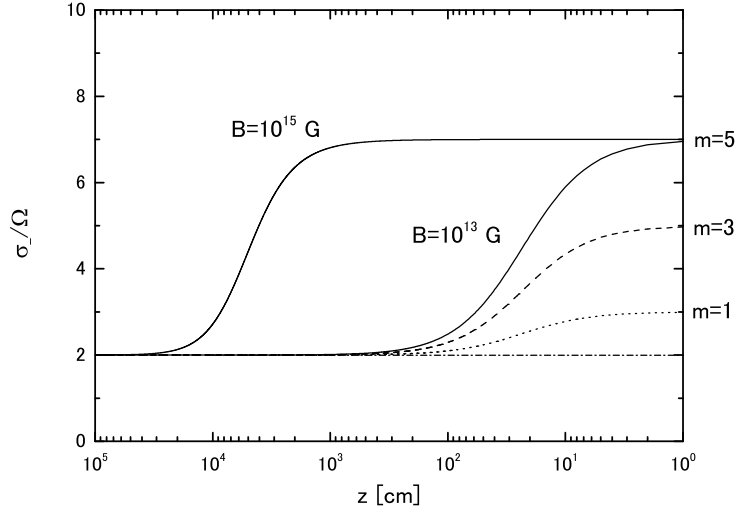}
\caption{ 
\small
The critical angular frequency $ \sigma_{-} /\Omega = 2 + m (1-h) $  
as a function of the depth from the surface.
Evanescent region for negative helicity mode $ \xi_{-}$ 
is given by the condition
$ 0 < \sigma _{-}/\Omega  < 2 + m (1-h) $.
The critical frequencies are shown for 
$ m=1,3,5$ with $B_0=10^{13}$G and $m=5$ with $B_0=10^{15}$G.
Note that the evanescent condition becomes 
$0< \sigma _{-}/\Omega  < 2 $
denoted by a dash-dotted line,  
if all relativistic effects are neglected.
The evanescent region near the surface enlarges with 
increase of $m$ and the magnetic field strength.
}
\end{figure}

%
The phase and group velocities of the shear-\ALF waves 
in the inertial frame are respectively given by
$v_p= \omega/k$
and
$v_g= \partial \omega/ \partial k$.
The following relation is easily calculated
\begin{equation}
 \frac{ v_p v_g }{v^2}  = 
\frac{ \omega }{
\omega \pm \frac{1}{2}
[ 2 -m(1+h) ]  \Omega }
=  \frac{ \sigma _{\pm}  \pm m \Omega}{
\sigma _{\pm}  \pm \frac{1}{2}
[ 2+ m(1-h) ]  \Omega }.
\end{equation}
This shows an interesting property that 
the phase and group velocities of the mode $\xi_{-} $
are opposite in the propagating direction 
for the frequency range
$ [ 2 +m(1-h)  ]  \Omega  < \sigma_{-}  <m  \Omega $,
which is possible for $m \ge 3$.
The lower bound comes from the condition for the 
wave propagation.

   In Fig.2, we show 
$ 1/N^2 = \omega ^2/(c^2 k^2)$ 
as a function of  $ \sigma_{-}/\Omega =\omega/\Omega + m$,
using the local dispersion relation (\ref{disp.eqn}).
The frequency corresponding to negative value of $ 1/N^2 $
mean evanescent.
The non-propagating frequency is located in low frequency region.
We now consider the propagation of the negative helicity mode
with $ 2 < \sigma_{-} /\Omega < m$.
The mode propagates as a wave in deep interior, and may
leak out through to the exterior. 
The amplitude of the outgoing wave becomes small
through the evanescent region. 
It is important to know the magnitude.
The damping rate depends on the frequency
and the depth $d_* $.  
The exponential damping factor $e^{- \kappa d_* }$ is
estimated as 
$ \kappa d_* \approx |N| \omega d_*/c $, which
is not so large as roughly estimated  $\sim 10^{-4}$ 
for large $m$.
This means that 
the damping of the amplitude is not so significant,
and therefore the disturbances of some modes
are penetrated into the surface.

\begin{figure}[ht]
\centering
\includegraphics[scale=1.0]{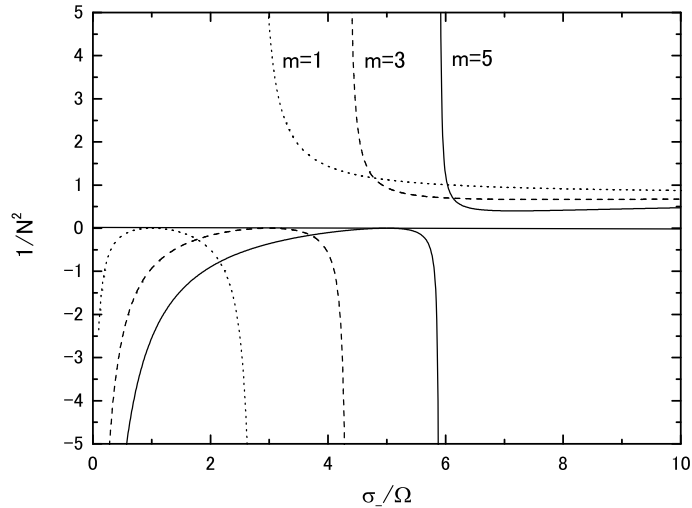}
\caption{ 
\small
Dispersion relation of negative helicity mode $ \xi_{-}$ 
as a function of angular frequency 
$ \sigma_{-}/\Omega $ at $z =10 $cm  from the surface.
}
\end{figure}
%

\subsection{ Exterior }

%
There are at least two possibilities as for the exterior of 
the stars. One is that surrounding plasma outside the star
corotates with the same angular velocity.
When the plasma is frozen in closed magnetic field lines, 
such a corotation is realized.  However, the plasma freely 
moves to infinity along open field lines, so that
the angular velocity is arbitrary in this case.
We also consider this situation by assuming that
the  plasma exterior to the star is
static in the inertial frame.
The difference of the plasma motion in the
background state is not so important 
for the high frequency modes of the perturbations
$ (\omega \gg \Omega )$,
but crucial for the low frequency modes, as discussed below. 

When the plasma corotates with the same angular velocity
$\Omega$, the perturbation equation
corresponds to the limit of eq.(\ref{base.eqn}) with
$ \rho = \mu =0$.
The wave solution is 
possible for the modes with
$ (\omega \mp m \Omega) (\omega \pm 2 \Omega)>0 $ 
and outgoing wave is expressed as
\begin{equation}
  \xi _{\pm}= A_{\pm} \exp[ i k_{\pm} z  ] ,
\label{extsl}
\end{equation}
where
\begin{equation}
  k_{\pm} =  
\frac{  \left[  (\omega \mp m \Omega) (\omega \pm 2 \Omega) 
        \right]^{1/2} }{c}  .
\end{equation}
We assume that the interior and exterior solutions
are continuously matched, so that 
the boundary condition of the displacement 
at $z=0$ can be expressed as
\begin{equation}
\label{bcrt}
  \frac{ d \xi_{\pm} }{d z } = i k_{\pm}\xi_{\pm} .
\end{equation}
We next consider the non-rotating plasma in the exterior of
the star. In this case, outgoing boundary condition at the 
surface is replaced as
\begin{equation}
\label{bcst}
  \frac{ d \xi_{\pm}}{d z } = i \frac{\omega} {c} \xi_{\pm} .
\end{equation}
  The shaking of the magnetic field lines induces
electromagnetic radiation in the exterior vacuum.
Since $\delta B_z =0$ everywhere,
the perturbation should be matched with the transverse
magnetic mode with the cylindrical wave guide
\citep{jac}.
The perpendicular components of electric fields,
i.e. $\delta E_\varpi , \delta E_\phi $ are continuous.
They are expressed by $\VXI$ in the interior and
vary as $ \propto \exp \{i \omega (z/c -t) \}$ in the exterior.
Thus we have
the boundary condition (\ref{bcst}) at the surface.
  
\subsection { Transmission coefficient }

In this section, we consider how much  \ALF waves excited 
in the crust are transmitted to the exterior.
We numerically solve eq.(\ref{base.eqn}) from the deep interior
$z\approx 10^{5}$ cm, corresponding to the neutron drip in the density, 
to the surface.
The following asymptotic solution is obtained
in the deep crust \citep{BBGP}. It is a sum of 
outgoing and reflecting waves, which respectively
propagate to $+z$ and $-z$ directions.
We normalize the amplitude of outgoing wave to unity 
for simplicity and denote that of reflecting wave
by  $R_{\pm}$.
\begin{equation}
\xi _{\pm }\approx \left| z\right| ^{-\frac{7}{4}}\left\{ \exp 
\left[ i\int \frac{\Gamma _{\pm }}{v} dz \right] 
+R_{\pm}\exp \left[ -i\int \frac{\Gamma _{\pm }}{v} dz \right] \right\} . 
\end{equation}
Here, $\Gamma _{\pm }$ is an effective
frequency of each wave mode defined as
\begin{equation}
\Gamma _{\pm }= \{
\left( \omega \mp m\Omega \right) 
\left[ \omega \mp \left(m h -2\right) \Omega \right] \}^{1/2}.  
\end{equation}
The outgoing wave solution in the
exterior is  given by eq.(\ref{extsl})
with a transmission amplitude $A_{\pm }$.
We here assume that the exterior plasma corotates with $\Omega$.
The interior solution should be continuously matched
with the exterior one at the surface, $z=0$.
Thus, the transmission coefficients 
$T_{\pm}$ for each helicity mode are given by the relation
\begin{equation}
T_{\pm }=\left| A_{\pm }\right| ^{2}=1-\left| R_{\pm}\right| ^{2}.  
\end{equation}
Our numerical calculation is limited to  $m=1$.
The lower bound of the frequency for the positive helicity mode
is $\sigma _{+} >0$, that is  $ \omega  > \Omega$.
On the other hand, that for negative helicity mode
is limited by propagation condition 
$\sigma _{-} > 3\Omega $, that is  $ \omega  > 2 \Omega$.
In Fig.3, transmission coefficients $T_\pm$ 
for each helicity mode are shown as a function 
of angular frequency in inertial frame $\omega$.
The rotational effect may be important in 
the millisecond magnetar model.
 In the calculation, the star
is assumed to rotate with a constant angular velocity, 
$\Omega /(2\pi) = 10^{3}$ Hz. 
This value is somewhat too rapid for SGRs, but we can regard it as 
an extreme one, which may be present at the newly borne phase.
The rotational effect becomes small for much smaller value
of $\Omega $.
For a comparison, we also show the result $T_0$ for 
the non-rotating case ($\Omega =0$) by solid curves in Fig.3.
The difference between positive and negative helicity states
is clear near the low frequency limit, i.e.  $\omega <$
a few $ \times \Omega $.
The difference however becomes almost
negligible in the high frequency regime, say, $\omega >10\Omega $.
The transmission rates gradually approach unity with increase of
the frequency.
The rates also increase with magnetic field strength,
but $ T_{\pm} \ne 1$ in the low frequency regime 
even for $10^{15}$ G. 
A significant fraction of waves suffers from the bounce at
the surface, because of $ T_{\pm} \sim 0.1$ for
$ \omega < 10\Omega $.
This means that the transmission time to the exterior
is roughly estimated not by the crossing time $ d/v $, 
but by $ d/v \times T_{\pm} ^{-1} $.

\begin{figure}[ht]
\centering
\includegraphics[scale=1.0]{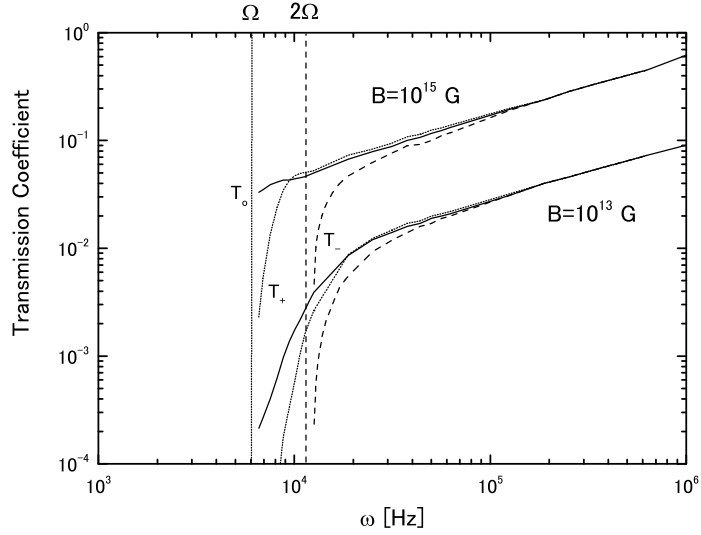}
\caption{ 
\small
Transmission rates as a function of angular frequency 
for the magnetic field strength $10^{13} $ and $10^{15} $G. 
The curves with labels $ T_+ $ and $ T_-$
denote results for the positive and negative helicity states.
The solid curves with the label $ T_o $
denote results for the non-rotating star.
}
\end{figure}
%

\subsection {Energy and angular momentum loss }

  We will calculate the rate of energy and angular momentum 
carried by the \ALF waves. 
The terms linear to the perturbation quantities vanish 
in the time average. 
Meaningful time-averaged values  come from
square of the  perturbation quantities.
The time-averaged Poynting flux is therefore given by
\begin{equation}
 \vec{S} = \frac{c}{8\pi} \Re \left(
\delta \EV \times \delta  \BV ^* \right) ,
\end{equation}
where $^* $ means complex conjugate
and $\Re() $ means taking the real part.
We have here used a useful technique in averaging 
the product of complex quantities with
the same  harmonics time-dependence
as shown in the textbooks, e.g. \cite{jac}.
The power  $P$  radiated across the surface 
at $z=0$  is calculated by integrating $S_z$
over a circle within $ \varpi_* $ as
\[
P  = \frac{c}{8\pi} \int _0 ^{\varpi_*} \int _0 ^{2 \pi}
\Re \left( 
\delta E_\varpi ^*   \delta B_\phi -
\delta E_\phi ^*     \delta B_\varpi \right) \varpi d\varpi d \phi
\]
\begin{equation}
~~= \frac{(\omega \mp m \Omega)  B_0 ^2\varpi^{2m} _* }{4m }
  \Re \left( -i  \xi^* _{\pm}  
  \frac{d\xi _{\pm}}{dz}  \right)_{z=0}. 
\label{Ez}
\end{equation}
This is roughly estimated as
$P \sim (\delta B)^2 ( \pi \varpi_* ^{2} ) c$.
The power $P_{(R)}$ ejecting to the 
corotating ambient plasma
is evaluated using the boundary condition (\ref{bcrt}).
The power $P_{(S)}$  to the static plasma
is also calculated with condition (\ref{bcst}).
The explicit expressions are 
\begin{equation}
P_{(R)}  = 
\frac{(\omega \mp m \Omega)  k_{\pm} 
B_0 ^2\varpi^{2m} _*  |\xi_{\pm}|^2 _{z=0}  }{4m },
\end{equation}
\begin{equation}
P_{(S)} = 
\frac{(\omega \mp m \Omega) \omega
B_0 ^2\varpi^{2m} _*  |\xi_{\pm}|^2 _{z=0} }{4m c} .
\end{equation}

In a similar way, 
the angular momentum flux per unit time across
$z=0$ can be calculated.
The transport of the angular momentum 
results in the torque.
The time-averaged torque due to the electromagnetic stresses
can be expressed in terms of the surface integral.
From the symmetry, non-vanishing  angular momentum flux
is $z$-component only, which is expressed as
\[
N_z  = - \frac{1}{8\pi} \int _0 ^{\varpi_*} \int _0 ^{2 \pi}
\Re \left( 
\delta B_\phi ^* \delta B_z +
\delta E_\phi ^* \delta E_z \right) 
\varpi^2 d\varpi d \phi
\]
\begin{equation}
~~~= \frac{(\omega \mp m \Omega)
  \Omega B_0 ^2 \varpi^{2m+2} _*  } { 8(m+1)c^2}
 \Re \left( -i  \xi ^* _{\pm}  \frac{d\xi _{\pm}}{dz}
 \right)_{z=0} .
\label{Jz}
\end{equation}
The angular momentum transfer in eq.(\ref{Jz})
comes from the electric part only, since 
$\delta B_z =0$.
The value is estimated as  
$N_z \sim (\Omega \varpi_*/c ) \times
(\delta B)^2 \varpi_* ( \pi \varpi_*^{2} )$,
and is smaller by an extra factor 
$ (\Omega \varpi _* /c ) $ 
compared with the rough estimate by the magnetic part only.
Note that the angular momentum transfer 
comes from the magnetic term, i.e. $B_\phi B_z $
for non-relativistic case such as interstellar clouds. 
Depending on the boundary condition (\ref{bcrt}) or (\ref{bcst}),
the angular momentum flux is written as
\begin{equation}
N_{z (R)}  = 
\frac{(\omega \mp m \Omega)  k_{\pm} \Omega
B_0 ^2\varpi^{2m+2} _*  |\xi_{\pm}|^2 _{z=0} }
{ 8(m+1)c^2} ,
\end{equation}
\begin{equation}
\label{NZST}
N_{z (S)} = 
\frac{(\omega \mp m \Omega) \omega \Omega
B_0 ^2\varpi^{2m+2} _*  |\xi_{\pm}|^2 _{z=0} }{ 8(m+1)c^3} .
\end{equation}
It should be noted that
the angular momentum flux can be negative
by the outgoing waves into the static 
ambient plasma as shown in eq.(\ref{NZST}).
This is possible if the condition 
$ (\omega \mp m \Omega) \omega  = \sigma _{\pm} \omega <0 $ 
is satisfied.
That is, frequencies, $\sigma $ and $\omega $,
are opposite in sign.
This condition is the same as the criterion of 
gravitational radiation reaction instability 
\citep{frsc78}.
The oscillation mode counter-rotates
when viewed in rotating frame with star,
but corotates when viewed in the inertial frame.
This provides a mechanism for converting the stellar 
rotational energy into gravitational radiation.
The modes satisfying the condition  
$ (\omega \mp m \Omega) \omega  <0 $ 
are very interesting. They carry negative
angular momentum.
The emission leads to the {\it  spin-up } of the star.
%

\subsection { Order of magnitude }

  There is a simple relation between the 
energy flux (\ref{Ez})  and  angular momentum flux (\ref{Jz}) 
exerted by  the \ALF waves: 
\begin{equation}
\label{NPrel}
N_z =  \frac{m \Omega \varpi^2 _* }{ 2(m+1) c^2 }   P .
\end{equation} 
This holds irrespective of the boundary condition
(\ref{bcrt}) or (\ref{bcst}).
The total angular momentum loss/gain
is related with the total energy radiated by the waves.
During the waves emission,
the stellar structure may change due to
re-arrangement of magnetic fields,
and the inertial moment $I$ may be modified.
For giant bursts in SGRs, such a possibility
is discussed \citep{io,w0}.
We do not consider the  possibility here.
The duration time of the wave emission 
is denoted by $\tau$.
The change in the angular velocity
$\Delta \Omega$ is expressed  by
the total energy $\Delta E_A= P \tau $
\begin{equation}
 \frac{\Delta \Omega }{\Omega} =\frac{N_z \tau }{I\Omega} 
 =  \frac{m \varpi^2 _* }{ 2(m+1)c^2 I } \Delta E_A
%
 \sim   10^{-20} 
\left( \frac{\varpi_*}{10^4 {\rm cm} }\right) ^2
\left( \frac{\Delta E_A}{10^{38} {\rm ergs} } \right),
\end{equation}
where the total energy by the \ALF waves
is not known, but is used for the typical burst
energy in observed in SGRs.
There are reports of glitches observed in 
AXPs\citep{ka},
$\Delta \Omega / \Omega \sim 10^{-6} $.
Our estimate is too small.
Much larger total energy may be
expected associated with the giant bursts.
Even taking comparable energy 
observed in X-/$\gamma$-rays in such events
$ \Delta E_A \sim 10^{44}$ergs,
the change is $\Delta \Omega / \Omega \sim 10^{-14} $,
whereas $|\Delta \Omega / \Omega | \sim 10^{-4} $ in 
the giant burst.
One reason of such smallness may come
from too much idealized situation, 
in particular, our model giving $ \delta B_z =0$ exactly.
This introduces a small factor
$(\Omega \varpi_*/c ) \sim 10^{-6}$
in the angular momentum loss $N_z$.
Moreover, larger size $\varpi_*$ should be required also. 
Otherwise, 
a wind of particles and MHD waves or
interaction between the crustal neutron superfluid
and the rest of the neutron star are more efficient 
processes for the angular momentum transfer \citep{tet1}.

\section { Discussion }

  We have studied the propagation of shear-\ALF waves
through the neutron-star crust to the exterior.
Using a simplified cylindrical model, 
some interesting natures are found.
The waves with  high frequency, $|\omega /( m \Omega)| \gg 1$,
always propagate,
whereas those with low frequency 
propagate or non-propagate depending on spatial position.
The transmission coefficient, which depends 
on the helicity state, is explicitly calculated.
The ejection rate to the exterior increases with the frequency 
of the wave and the magnetic field strength.
  We now consider the behavior of  modes in low frequency regime.
The positive helicity modes $\xi _{+}$ always propagate
as far as  $\sigma_{+} >0   $, whereas
the negative helicity modes $\xi _{-}$ become
evanescent near the surface for 
$ 0 < \sigma _{-} < (m+2)\Omega $.
If corotating plasma covers the surface, 
the disturbances do not propagate in the atmosphere.
On the other hand, if such plasma does not exist,
i.e. static plasma or vacuum case,
then the perturbations revive as outward waves in the exterior, 
although they suffer from some damping through the interior 
evanescent region.
The exponential decay of the amplitude is not so large,
for $ \sigma _{-} \sim m \Omega $ with large $m$, 
as shown in section 2.3.
The disturbances therefore penetrate into the surface, and
eventually escape to infinity.

When the waves with the frequency $0<  \sigma _{-} < m \Omega$
are ejected, they carry away
negative angular momentum through the surface.
In this case, the torque  becomes positive
and leads to  {\it  spin-up } of the star
by the outgoing waves.
However, the magnitude is not enough to explain
the glitches observed in AXPs\citep{ka}.
The geometrical factor may be important in  actual situation.
In our model, the spin axis agrees with
the direction of wave propagation,
and the direction of the perturbations is 
orthogonal to the spin axis.
For magnetic field line mis-aligned with the rotation axis,
the direction of the perturbations is perpendicular to
the propagation, but not to the spin axis. 
In this case, $\delta B_z$ is induced in general, and the
non-vanishing term $\delta B_\phi \delta B_z$ 
results in larger torque.  
See eq.(\ref{Jz}) and the discussion.
Compressional fast/slow magnetosonic waves are also induced
at the same time.
The calculations would be much more complicated.
   The increase of angular momentum as
a result of radiation is very analogous to
that in  the gravitational radiation reaction 
instability, i.e. the CFS mechanism
\citep{ch70,frsc78}.
The generic criterion of the instability 
is  $ \sigma_{\pm} \omega < 0 $
for unmagnetized rotating stars.
Some fluid oscillation-modes satisfying this condition
can grow in secular time scale 
as a reaction of gravitational radiation.
The low frequency mode satisfies the instability criterion.
Related with this, there is a suggestive work by \cite{hl}. 
They considered the r-mode instability driven by
\ALF wave emission as well as gravitational one.
On the dimensional estimate,
the transfer rate by \ALF wave is
larger than that by  gravitational wave,
for highly magnetized star 
$ B \ge 3 \times 10^{12} (T/10{\rm ms})^{-3}$G.
They pointed out the \ALF wave-driven instability as
an efficient process.
In their treatment, however, the displacement current 
is neglected, and hence dynamical degree of the
electromagnetic fields is eliminated in a sense.
Such an approximation may be justified in the deep 
interior, but no longer near the surface.
Our treatment is concentrated on the crust, but
not the interior core, whose bulk motion is
important to estimate gravitational radiation.
Thus global calculation adequate for the emission
of gravitational wave and \ALF wave is required
to determine the efficient mechanism.
As for a fundamental problem of the secular instability,
the  criterion should be clarified by
taking into account of strong  magnetic fields.
For example, the canonical energy relevant to the
instability is constructed only for unmagnetized stars \citep{frsc78}.
In their treatment of the secular instability,
the effect of gravitational emission is
neglected in dynamical equation, but
should be  accounted in the energy equation governing 
the evolution in longer time scale.
Our problem differs in this point.
The  degree of freedom of the electromagnetic 
fields is accounted in dynamical equation,
and hence \ALF wave emission from surface
is inevitably included.
There is no mathematically rigorous proof
of the secular stability criterion in magnetized stars,
so that further study is necessary to conclude unstable
growth of the modes coupled with electromagnetic fields.

Finally, we will comment on some problems in applying to
more realistic models relevant to magnetars.
The rotation of observed magnetars is very slow
$\Omega \sim 1$rad s$^{-1}$.
In order to satisfy the condition  $ 0< \sigma_{-}  < m  \Omega $,    
a large azimuthal number, say,  $m \sim 10^3$ is needed.
If there are rapidly rotating stars,
a moderate value of $m$ may be sufficient.
The other important factor is the magnetic configuration. 
\cite{tlk} considered the magnetosphere
threaded by large-scale electric current.
The  magnetic field structure is twisted, and
quite different from our present model.
The perturbations of such magnetic fields are
induced by bursts, and \ALF waves propagate.
However, the detailed treatment seems to be  complicated,
and highly numerical calculations are needed.
We expect that the behavior of short wavelength,
say, much smaller than curvature of magnetic field,
may be almost the same as given here.
The result of the transmission rate calculated 
in section 2.5 may be relevant.
For the long wavelength mode, 
geometrical effects are important, and
further studies are required.
%

\section*{Acknowledgements}
This work was supported in part 
by the Grant-in-Aid for Scientific Research 
(No.14047215, No.16029207 and No.16540256) from 
the Japanese Ministry of Education, Culture, Sports,
Science and Technology.


%
\end{document}